\documentclass[,floats,superscriptaddress]{revtex4}

\usepackage[dvips]{graphicx}
\usepackage{amssymb}
\usepackage{amsmath,bm}
\usepackage{psfrag}
\usepackage{epsfig}
\usepackage{float}

\begin{document}

%*******************************************************************
%TITLE OF THE ARTICLE
%*******************************************************************
\vspace*{2cm} \normalsize \centerline{\Large \bf Electromigration-driven Evolution of the Surface Morphology }
\medskip
\centerline{\Large \bf and Composition for a Bi-Component Solid Film}%\vspace*{0.2cm}\\

\vspace*{1cm}
%*******************************************************************
%AUTHORS - THE CORRESPONDING AUTHOR NEEDS TO SPECIFY HIS/HER E-MAIL ADDRESS AS A FOOTNOTE
%*******************************************************************

\centerline{\bf Mikhail Khenner$^{a}$, Mahdi Bandegi$^b$}

\vspace*{0.5cm}

%*******************************************************************
%ADDRESS OF THE AUTHORS
%*******************************************************************
\centerline{$^a$ Department of Mathematics and Applied Physics Institute,} 
\centerline{Western Kentucky University, Bowling Green, KY 42101} 
\centerline{E-mail: mikhail.khenner@wku.edu}

\centerline{$^b$ Department of Mathematics,}
\centerline{Western Kentucky University, Bowling Green, KY 42101}
\centerline{Current affiliation: Ph.D. program, Department of Mathematical Sciences,}
\centerline{New Jersey Institute of Technology, Newark, NJ 07102}

%*******************************************************************
%ABSTRACT
%*******************************************************************

%\vspace*{1cm}

\begin{abstract}
A two PDEs-based model is developed for studies of a morphological and compositional evolution of a thermodynamically stable alloy surface in a strong electric 
field, assuming different and anisotropic diffusional mobilities of the two atomic components. The linear stability analysis of a planar surface and the computations of 
morphology coarsening are performed.
It is shown that the conditions for instability and the characteristic wavelength and growth rate differ from their counterparts in a single-component film.
Computational parametric analyses reveal the sensitivity of the scaling exponents to the electric field strength and to the magnitude of the anisotropies difference.


\end{abstract}

\begin{center}
\small{{\it Journal information: Mathematical Modelling of Natural Phenomena Vol. 10, No. 4, 2015, pp.83-95; DOI: 10.1051/mmnp/201510405}}
\end{center}

%*******************************************************************
%NECESSARY TO OBTAIN THE SECTIONS, SUBSECTIONS, ETC. DEFINED AS 1.1., 2.1.3. FOR INSTANCE
%*******************************************************************

\setcounter{equation}{0}
\section{Introduction}
\label{Intro}

Surface electromigration \footnote{Defined as the drift, usually in the direction of DC electric current,  of the ionized adsorbed atoms (adatoms) due to their interaction with the "electron wind".} was  studied theoretically 
in connection to the grain-boundary grooving in polycrystalline films \cite{M}-\cite{AO}, the kinetic instabilities of crystal steps \cite{St}-\cite{UCS}, morphological 
stability of thin films \cite{KD}-\cite{K}, and recently, as a way to fabricate nanometer-sized gaps in metallic films - suitable for testing of the conductive properties of single molecules and control of their functionalities \cite{VFDMSKBM}-\cite{GSPBT}.

This paper theoretically investigates the effects of electromigration on morphological stability and evolution of a bi-component, atomically rough surface of a single-crystal metal or semiconductor film 
(a generic substitutional binary alloy or a non-reactive compound). The prototype systems may be the Al$_x$Cu$_{1-x}$ or Ag$_x$Pd$_{1-x}$ grain of a microelectronic interconnect, or a Si$_x$Ge$_{1-x}$ thin film. Here $0<x<1$ stands for the concentration of Al, Ag or Si atoms. The electric field is assumed applied along
the direction of the initially planar grain or film surface (or parallel to a crystal step in the step dynamical setting), and the thin surface layer contains both atomic components. Such direction of the application of the electric field  
induces faceting of the film surface, which was the subject of several studies of a single-component films \cite{DDF},\cite{KD}-\cite{BMOPL},\cite{ZYDZ}.

In microelectronic applications, the performance and realiability of alloy interconnects depends in part on the distribution of the minority component \cite{CR,HSKM}. The electric current and associated heating of the 
interconnect may affect the minority component distribution within the surface layer and in the bulk of the grain. Often a new phase(s) is formed, which occurs primarily at the grain
boundaries. As interconnects dimensions approach the nanoscale, their resistance to electromigration-caused degradation is expected to diminish \cite{VFDMSKBM}-\cite{GSPBT},\cite{ZYDZ}, thus understanding 
how film morphology is affected by electromigration is still important. Also, despite the apparent importance of a second atomic component, there have been no attempts to theoretically understand how its electromigration-driven surface diffusion 
affects the spatio-temporal distribution of the alloy components and the evolution of surface morphology. This paper is aimed to partially fill this gap.

\setcounter{equation}{0}
\section{Problem statement}
\label{ProbStat}

\begin{figure}[h!]
\vspace{-4.5cm}
\centering
\includegraphics[width=4.5in]{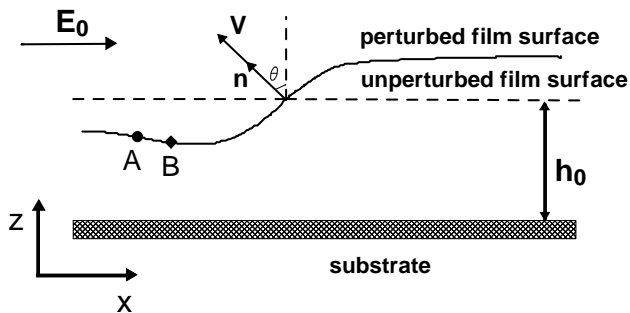}
\vspace{-5.5cm}
\caption{Sketch of the problem geometry. }
\label{Geom}
\end{figure}

We assume a simple one-dimensional geometry, where the surface is an open curve (without overhangs) in the $xz$ plane, described by a function $z=h(x,t)$.  The surface diffusional mobilities $M_A$ and $M_B$,  where $A$ and $B$ are two types of mobile adatoms, are usually anisotropic due to the undelying crystal lattice \cite{KD,SK}.  
Thus $M_i=M_i(\theta(s))$, $i=A, B$, where $\theta$ is the surface orientation angle and the arclength $s$ is the position variable (see Figure \ref{Geom}).
As was noted, our local model assumes that the constant electric field vector $\mathbf{E_0}$ is directed along the substrate.
The component of $\mathbf{E_0}$ that is parallel to the surface is the origin of the electromigration force on adatoms. Thus the electromigration force is a function of the surface orientation angle $\theta$; and since $\theta$ changes from point to point on the surface, it follows that the force depends on the
arclength $s$ (and thus it depends on $x$).
We also assume:
\begin{itemize}
\item The post-deposition scenario, when the surface shape changes by the natural, high-temperature surface diffusion of adatoms 
(which arises due to a non-uniformity of the surface chemical potential $\mu$ along the surface), 
and by the electromigration-driven surface diffusion \cite{M};
\item  Surface orientation independent (isotropic) and composition independent surface energies $\gamma_i$ (thus $\gamma_i=const.$). Typical solid films feature anisotropic and (weakly) composition-dependent surface energies $\gamma_i(\theta,C_i)$, but in the presence of the electric field (which sets up a preferred direction for adatoms diffusion)
it is expected that these effects are less important than the effects caused by the anisotropy of the diffusional mobilities. Morphological evolution with the anisotropic surface energy 
has been extensively studied, see for instance \cite{LM,SGDNV};
\item Equal, i.e. the same value and sign, effective charges of $A$ and $B$-type adatoms.
This leads to the simpler forms of the governing equations and the reduction from six to five in the number of independent parameters of the dimensionless problem. 
Even when the effective charges are of the same sign, as is generally expected, it is quite reasonable to assume that their values ratio
%differ by as much as a hundred times.
may be as large as 100.
Thus we will report separately on the results of the analysis of the full problem where this assumption is relaxed.
\end{itemize}
Furthermore, the stresses in the film are ignored; this includes thermal, compositional and heteroepitaxial stresses.

Let $C_A(x,t)$ and $C_B(x,t)$ be the dimensionless surface concentrations of adatoms $A$ and $B$, defined as the products of a volumetric number densities and the atomic volume. 
Then $C_A(x,t)+C_B(x,t)=1$, and it is sufficient to determine the concentration of one adatom type, say, $C_B(x,t)$.
%
%\begin{figure}[h!]

The model that we develop is aimed at the description and understanding of the conditions leading to destabilization of the initially planar surface and the time-evolution of the surface shape and composition after the destabilization occured. 
The surface shape $h(x,t)$ and the concentration $C_B(x,t)$ will be determined from an initial-boundary value problem for a system of two coupled, well-posed parabolic partial differential equations (PDEs).
To this end, our model is largely rooted in the theoretical framework developed by Spencer, Voorhees, and Tersoff \cite{Spencer_concentration} 
for the analysis of the morphological evolution of the surfaces of a bi-component, heteroepitaxial thin films. The major attraction of this model is that each component is allowed to
diffuse independently on the surface, which permits to determine the impact of each component diffusional mobility (and its anisotropy). As was already mentioned, this anisotropy is the important factor in surface 
electromigration phenomena \cite{KD}-\cite{BMOPL},\cite{TGM1,ZYDZ}.

From geometry, the PDE for $h(x,t)$ reads: 
\[
h_t=V/\cos{\theta},
\]
where following \cite{Spencer_concentration} %and \cite{Klinger} 
the normal velocity of the surface is given by:
\[
V = -\Omega\left(\frac{\partial J_A}{\partial s}+\frac{\partial J_B}{\partial s}\right).
\]
Here $J_A$ and $J_B$ are the surface diffusion fluxes of the components $A$ and $B$. (Most physical parameters are described in the Table \ref{table1}, thus in the text
we only give descriptions of the parameters that are not in that Table, or that need clarifications.)

Accounting for the two contributions to the surface diffusion in a usual way, i.e. using the Nernst-Einstein relation, gives the expressions for the fluxes \cite{TGM1}:
\begin{equation}
J_i = -\frac{\nu D_i}{kT}M_i(\theta)C_i(s,t)\left[\frac{\partial \mu_i}{\partial s} + \alpha E_0 q \cos{\theta}\right].
\label{J-eq}
\end{equation}
%where $C_i(s,t)$ is the fraction of adatoms which is component $i$, and $E_0$ is the electric field strength.
Here $E_0\cos{\theta}$ is the component of $\mathbf{E_0}$ parallel to the surface. The local approximation for the electric field in Eq. (\ref{J-eq}) 
%loses precision
becomes less accurate
when the deviations from the planar surface morphology become large. This can be corrected by solving the boundary-value problem for the electric potential in the bulk of the solid 
and then using the solution value on the surface to compute the electromigration flux. Solution of the full electrostatic problem can be accomplished numerically at every step 
of the time-marching method for the surface evolution PDE, see for instance \cite{MB,KAING,SK,TGM1}. Due to using the local approximation in Eq. (\ref{J-eq}), the evolution equations that we derive next do not require the bulk electrostatic potential field. We choose to use the local approximation since (i) it facilitates the stability analysis in Sec. 3, and (ii) 
the accurate numerical solution of the electrostatic problem is expected to be challenging, at least in the beginning of the simulation, for the pointwise random initial conditions that are desirable in the studies of the morphology coarsening in Sec. \ref{Computations}
\footnote{The examples of such initial conditions are shown in Figures \ref{h_CB_beta_A=0.1} and \ref{h_CB_beta_B=0.1}. Although Schimschak and Krug \cite{SK} implemented the boundary-value problem solver for the case of random initial surface profiles in a one-component system, they did not report the numerical convergence results, and they did not extract the coarsening exponents from their computations, instead focusing on the effect of the lateral drift of the surface perturbations and on the late 
time steady states. 
%The high accuracy of the numerical scheme at the beginning of the random-initial condition computation is very important, in our opinion, as it may at large 
%determine the overall acccuracy of the coarsening exponents.
}

We choose to express the diffusional mobilities as in \cite{SK}:
\begin{equation}
M_i(\theta)= \frac{1+\beta_i\cos^2{[N_i(\theta+\phi_i)]}}{1+\beta_i\cos^2{[N_i\phi_i]}}, \label{Mobility}
\end{equation}
where $\beta_i$ is anisotropy strength, $N_i$ is the number of symmetry axes, and $\phi_i$ is the angle between a symmetry direction and the average surface orientation.
Eq. (\ref{Mobility}) is dimensionless.

Evolution of the surface concentration $C_B(s,t)$ is governed by the PDE (see Appendix A in \cite{Spencer_concentration}):
\[
\delta \frac{\partial C_B}{\partial t} + C_B V =  -\Omega\frac{\partial J_B}{\partial s},
%\label{C-eq}
\]
where $\delta$ is the thickness of the surface layer and quantifies the ``coverage". 

Finally, the surface chemical potentials of the components are given by:
\[
\mu_i = \Omega \gamma_i\kappa + \mu_i^0\left(C_i\right),
%\label{1.4e}
\]
where $\kappa = \partial \theta/\partial s$ is the curvature, and
$\mu_i^0\left(C_i\right)$ are the thermodynamic contributions to the chemical potentials, written using the regular solution model of the mixture as \cite{ThermBook,Walgraef}
\begin{equation}
\mu_i^0\left(C_i\right) = kT ln\frac{C_i}{1-C_i}.
\label{mu_c}
\end{equation}
%where $kT$ is Boltzmann's factor.
We linearize $\mu_i^0\left(C_i\right)$ about the reference concentration $C_i=1/2$ \cite{Spencer_concentration} and obtain
\begin{equation}
\mu_i \approx \Omega \gamma_i\kappa -2kT +4kT C_i.
\label{base_eq4}
\end{equation}
Note that Eq. (\ref{mu_c}) implies a thermodynamically stable alloy, thus the natural surface diffusion  acts to smooth out any compositional nonuniformities. On the other hand, the electromigration may be the cause of their emergence and development.

Next,  using $\partial/\partial s = (\cos{\theta})\partial/\partial x=
\left(1+h_x^2\right)^{-1/2}\partial/\partial x$ and $\theta=arctan\left(h_x\right)$, we obtain the following two evolution PDEs:
\[
h_t=-\Omega\left(\frac{\partial J_A}{\partial x}+\frac{\partial J_B}{\partial x}\right), %\label{base_eq1}
\]
\[
\delta \frac{\partial C_B}{\partial t} = -\left(1+h_x^2\right)^{-1/2}\left[C_B  h_t   +\Omega \frac{\partial J_B}{\partial x}\right],
%\label{base_eq3}
\]
where 
\[
J_i = \frac{-\nu D_i}{kT}M_i\left(h_x\right)C_i\left(1+h_x^2\right)^{-1/2}\left[\frac{\partial \mu_i}{\partial x} + \alpha E_0 q \right],\quad C_i = C_i(x,t),
%\label{base_eq2}
\] 
\begin{equation}
M_i\left(h_x\right)= \frac{1+\beta_i\cos^2{[N_i(arctan\left(h_x\right)+\phi_i)]}}{1+\beta_i\cos^2{[N_i\phi_i]}},
\label{base_eq6}
\end{equation}
and the curvature in the expression (\ref{base_eq4}) for the chemical potentials is
\[
\kappa = -h_{xx}\left(1+h_x^2\right)^{-3/2}.
%\label{1.4z}
\]

Finally, we choose the height $h_0$ of the as-deposited film as the length scale and $h_0^2/D_B$ as the time scale. Also we take 
\begin{itemize}
\item $E_0=\Delta V/L$, where $\Delta V$ is the applied potential difference and $L=nh_0$ is the lateral dimension of the film ($n>0$ is a parameter),
\item $\delta=m\Omega \nu$, where $m>0$ is a parameter. 
\end{itemize}
Then, the dimensionless system of coupled, highly nonlinear evolution PDEs takes the final form:
\begin{equation}
h_t=\frac{4}{mQ}\frac{\partial}{\partial x}\left\{\left(1+h_x^2\right)^{-1/2}\left[D M_A\left(h_x\right)(1-C_B)\left(R_A\frac{\partial \kappa}{\partial x}-
\frac{\partial C_B}{\partial x}+F\right)+\right.\right.
\label{nondim_h_eq2}
\end{equation}
\[
\left.\left.M_B\left(h_x\right)C_B\left(R_B\frac{\partial \kappa}{\partial x}+\frac{\partial C_B}{\partial x}+F\right)\right]\right\},
\]
\begin{equation}
\frac{\partial C_B}{\partial t} = -\left(1+h_x^2\right)^{-1/2}\left[QC_B  h_t -  \frac{4}{m}\right.
\label{nondim_C_eq2}
\end{equation}
\[
\left.\frac{\partial}{\partial x} \left\{\left(1+h_x^2\right)^{-1/2}M_B\left(h_x\right)C_B \left(R_B\frac{\partial \kappa}{\partial x}+\frac{\partial C_B}{\partial x} +F \right)\right\}   \right].
\] 
Here the parameters are:
\[
R_i=\frac{\Omega \gamma_i}{4kTh_0},\; F=\frac{\alpha \Delta V q}{4n kT},\;  Q=\frac{h_0}{m\Omega \nu},\; D = \frac{D_A}{D_B}.
\]
%Eqs. (\ref{nondim_h_eq2}) and (\ref{nondim_C_eq2}) are stated in their conservative forms. 
There is a total of five independent  parameters ($R_A,\;R_B,\;n,\;m,\;D$) 
since values of $F$ and $Q$ depend on $n$ and $m$, respectively.  $R_A$ and $R_B$ are the dimensionless surface energies of the components $A$ and $B$, $F$ is the
applied voltage parameter, $Q$ is the ratio of the film thickness to the thickness of the surface layer, and $D$ is the ratio of the diffusivities of the two components.
The first (second) term in the square bracket at the right-hand side of Eq. (\ref{nondim_h_eq2})
stands for the contribution of $A$ ($B$)-component. The first term is weighted by the diffusivities ratio, and the second term times the geometric factor $\left(1+h_x^2\right)^{-1/2}$ appears also at the right-hand side of Eq. (\ref{nondim_C_eq2}).
It can be seen that, if only the type $B$ adatoms are present (the standard case of a one-component film) and there is no applied potential difference, then $C_B(x,t)=1,\ F=0$; 
next, take $M_B\left(h_x\right)=1$ (isotropic diffusional mobility, $\beta_B=0$) and Eqs. (\ref{nondim_h_eq2}) and (\ref{nondim_C_eq2}) \emph{both} reduce to the same basic 
equation $h_t = (\Omega^2 \nu \gamma/kT h_0^2)\left[\left(1+h_x^2\right)^{-1/2}\kappa_x\right]_x$ first introduced by W.W. Mullins to describe the morphology evolution by surface diffusion \cite{M}. The reduction also makes it clear that the term $QC_B  h_t$ is necessary in Eq. (\ref{nondim_C_eq2}) even in the absence of the deposition flux \cite{Spencer_concentration}: when this term is omitted,
this equation becomes the static one $\left(1+h_x^2\right)^{-1/2}\kappa_x=const.$, which does not have a physical meaning. Alternatively, the limit of a one-component 
film is recovered when one sets $C_B=0$, $F=0$, and $DM_A\left(h_x\right)=1$ in Eqs. (\ref{nondim_h_eq2}) and (\ref{nondim_C_eq2}). Then the latter equation is 
identically zero (due only to vanishing $C_B$) and only the former equation transforms into the Mullins' equation cited above.

\setcounter{equation}{0}
\section{Linear Stability Analysis (LSA)}
\label{LSA}

We first linearize $M_i\left(h_x\right)$ about $h_x=0$, i.e. we write $M_i\left(h_x\right) = M_i(0)+M_i'(0)h_x$, where $M_i(0)$ and $M_i'(0)$ will be later calculated from Eq. (\ref{base_eq6}) for given 
$\beta_i,\ N_i$ and $\phi_i$ (see \cite{K}). Obviously, the derivative of the mobility with respect to $x$, which is needed in Eqs. (\ref{nondim_h_eq2}) and (\ref{nondim_C_eq2}), is
calculated using the Chain Rule as $\frac{\partial M_i\left(h_x\right)}{\partial x}= \frac{\partial M_i\left(h_x\right)}{\partial h_x}h_{xx} = M_i'(0)h_{xx}$.
%\begin{equation}
%\frac{\partial M_i\left(h_x\right)}{\partial x}= \frac{\partial M_i\left(h_x\right)}{\partial h_x}h_{xx} = M_i'(0)h_{xx}.
%\label{2.1}
%\end{equation}

Next, we take $h(x,t)=1+\xi(x,t), C_B(x,t)=C_B^0+\hat C_B(x,t)$
(where $\xi(x,t)$ and $\hat C_B(x,t)$ are \textit{small} perturbations) and linearize the PDEs. Then in these linear equations for
$\xi$ and $\hat C_B$ we assume $\xi(x,t) = Ue^{\omega(k) t}e^{ikx},\; \hat C_B(x,t) = Ve^{\omega(k) t}e^{ikx}$, where $U,\; V$ are the (unknown) constant and \emph{real-valued} amplitudes, 
$\omega= \omega^{(r)}(k) +i \omega^{(i)}(k)$ is the complex growth rate and $k$ is the wavenumber. This results in the algebraic system of two linear and homogeneous equations for the amplitudes $U,\; V$.   The matrix of this system is:
\[
\newline A=\frac{4}{m}\left (
  \begin {array} {cc} a_{11} & a_{12} \\ a_{21} & a_{22} \\\end {array} \right),
\]
where the complex elements are
\begin{eqnarray}
a_{11} &=& \frac{k}{Q}\left[-kM_B(0)C_B^0-kM_A(0)D \left(C_B^0-1\right)+iF\left(M_B(0)-D M_A(0)\right)\right], \nonumber\\
a_{12} &=& \frac{k^2}{Q}\left[k^2\left\{M_A(0)R_A D\left(C_B^0-1\right)-M_B(0)R_BC_B^0\right\} +F D M_A'(0)\left(C_B^0-1\right)- \right. \nonumber \\  
& & \left.FM_B'(0)C_B^0\right] - \omega, \nonumber \\
a_{21} &=& kM_B(0)\left(iF-kC_B^0\right) - \omega, \nonumber \\
a_{22} &=&  -\frac{m}{4}QC_B^0\omega - k^2C_B^0\left(FM_B'(0)+k^2R_BM_B(0)\right). \nonumber
\end{eqnarray}
Notice that the derivatives of the diffusional mobilities are slaved to the electric field parameter $F$, in other words the anisotropy
effect emerges only when the electromigration is operational \cite{KD,SK,K}.

A nontrivial solution of the algebraic system exists if and only if the determinants of the real and the imaginary parts of $A$ both equal zero. From the former condition one obtains
a quadratic equation for $\omega^{(r)}(k)$; its positive solution is the dispersion relation.
The equation reads:
\begin{equation}
\omega^{(r)}(k)^2 + \omega^{(r)}(k)\left[k^2 C_B^0\left(M_B(0)\left(1-\frac{m}{4}C_B^0\right)+\frac{M_A(0)}{4}mD\left(1-C_B^0\right)\right)\right. + \label{quadr_omega}
\end{equation}
\[
\left.\frac{k^2}{Q}\left\{DFM_A'(0)\left(1-C_B^0\right)+FM_B'(0)C_B^0+k^2DR_AM_A(0)\left(1-C_B^0\right)+k^2R_BM_B(0)C_B^0\right\}\right]+
\]
\[
\frac{k^4}{Q}\left[DF\left(M_B(0)M_A'(0)+M_A(0)M_B'(0)\right)C_B^0\left(1-C_B^0\right)+\right.
\]
\[
\left. k^2D\left(R_A+R_B\right)M_A(0)M_B(0)\left(1-C_B^0\right)\right] = 0.
\]

In the limit of a vanishing surface layer thickness, $\delta \rightarrow 0$ or equivalently, $Q\rightarrow \infty$ from Eq. (\ref{quadr_omega}) one obtains a simple expression
%at $m=1$,
\[
\omega^{(r)}(k) = -k^2 C_B^0\left[M_B(0)\left(1-\frac{C_B^0}{4}\right)+\frac{M_A(0)}{4}D\left(1-C_B^0\right)\right]. %\label{or_limitQ0}
\]
It can be seen that $\omega^{(r)}(k)<0$ (since the initial concentration $C_B^0 <1$), thus in this limit all perturbations 
decay.

\begin{figure}[h!]
\vspace{0.4cm}
\centering
\includegraphics[width=2.5in]{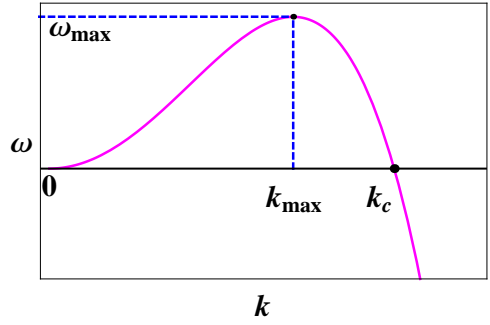}
%\vspace{-0.8cm}
\caption[\quad Sketch of the growth rate $\omega^{(r)}(k)$.]{(Color online.) Sketch of the growth rate $\omega^{(r)}(k)$ corresponding to a longwave instability of the film surface.}
\label{wk}
\end{figure}

A typical example of the dispersion curve for finite $Q$ is shown in Figure \ref{wk}. The surface is linearly unstable with respect to the long-wave perturbations
having wavenumbers $0<k<k_c$, where the cut-off wavenumber 
\begin{equation}
k_c = \left[F\left(1-C_B^0\right)\frac{M_B(0)M_A'(0)+M_A(0)M_B'(0)}{M_A(0)M_B(0)\left(C_B^0-1\right)\left(R_A+R_B\right)}\right]^{1/2} \label{kc}
\end{equation}
is the positive root of the equation $\omega^{(r)}(k)=0$.
%Notice that $k_c$ is independent of the diffusivities ratio.
The maximum perturbation growth rate is attained at $k=k_{max}$, where
$k_{max}$ is the root of the equation $d\omega^{(r)}(k)/dk=0$; correspondingly, $\lambda_{max}=2\pi/k_{max}$ is the wavelength of the most dangerous mode.
This mode will dominate over other modes shortly after the surface is destabilized, resulting (if one assumes vanishing lateral drift for a moment) in the surface deformation of the form $h(x,t)\approx 1 + a e^{\omega^{(r)}_{max} t} \cos{k_{max}x}$,
and the concentration $C_B(x,t) \approx C_B^0 + b e^{\omega^{(r)}_{max} t} \cos{k_{max}x}$, where $a,\ b \ll 1$ are the initial perturbations amplitudes, and 
$\omega^{(r)}_{max} = \omega^{(r)}\left(k_{max}\right)>0$.
Such exponential growth would continue until the evolution enters a nonlinear regime. 

\begin{table}[!ht]
\centering
{\scriptsize 
\begin{tabular}
{|c||c|c|p{4cm}|}

\hline
				 
			\rule[-2mm]{0mm}{6mm} \textbf{Physical Parameter}	 & \textbf{Typical Value} & \textbf{Fixed or Variable} & \textbf{Description}\\
			\hline\hline
			\rule[-2mm]{0mm}{6mm} $h_0$ & $ {1.0}\times{10^{-5}} cm$ & Fixed & Initial height of the film \\
			\hline
				\rule[-2mm]{0mm}{6mm} $\Omega$ & ${2.0}\times{10^{-23}} cm^3$ & Fixed & Adatom volume ($A$ or $B$ type)\\
			\hline
				\rule[-2mm]{0mm}{6mm} $\nu$ & ${1.0}\times{10^{15}} cm^{-2}$ & Fixed & Surface density of all ($A$ and $B$) adatoms\\
			\hline
			   \rule[-2mm]{0mm}{6mm} $D_A,\ D_B$ & ${{1.5}\times{10^{-6}}} {cm^2/s}$ & Variable & Surface diffusivity of $A$ or $B$ adatoms\\
			\hline
			\rule[-2mm]{0mm}{6mm} $\alpha$ & ${1}$ & Fixed & Sets the electric field orientation to result in long-wave surface instability\\
			\hline
				\rule[-2mm]{0mm}{6mm} $q$ & ${5.0}\times{10^{-11}} C$ & Fixed & Effective charge of $A$ or $B$ adatom\\
			\hline
			    \rule[-2mm]{0mm}{6mm} $\Delta V$ & $1$  V & Variable & Applied voltage\\
			%\hline
			    %\rule[-2mm]{0mm}{6mm} $L$ & ${1.0}\times{10^{-4}} cm$ & Fixed & Lateral extent of the film\\
			\hline
			   \rule[-2mm]{0mm}{6mm} $kT$ & ${1.12}\times{10^{-13}} erg$ & Fixed & Boltzmann's factor\\
			\hline
				\rule[-2mm]{0mm}{6mm} $\gamma _A,\ \gamma _B$ & ${2.0}\times{10^{3}} {erg/cm^2}$ & Fixed & Energy of a surface composed of $A$ or $B$ adatoms\\
			\hline
			    \rule[-2mm]{0mm}{6mm} $M_A(0),\ M_B(0)$ & 1 & Fixed & Diffusional mobility of $A$ or $B$ adatoms on the planar surface\\
			\hline
				\rule[-2mm]{0mm}{6mm} $M_A' (0),\ M_B' (0)$ & -2.67 & Variable & Derivative of the diffusional mobility of $A$ or $B$ adatoms on the planar surface\\
			\hline
			    \rule[-2mm]{0mm}{6mm} $C_A^0,\ C_B^0$ & 0.5 & Fixed & Initial fraction of $A$ or $B$ adatoms on the surface\\
                        \hline
				\rule[-2mm]{0mm}{6mm} $m$ & $1$ & Fixed & Coefficient in $\delta=m\Omega \nu$\\
                        \hline
				\rule[-2mm]{0mm}{6mm} $n$ & $10$ & Fixed & Coefficient in $L=nh_0$\\
			\hline				

\end{tabular}}
$$ $$
\caption[\quad Physical parameters]{Physical parameters. Cited values of $M_{A,B}'(0)$ correspond to $\beta_{A,B}=1,\ N_{A,B}=4,\ \phi_{A,B}=\pi/16$. 
($N_{A,B}$ and $\phi_{A,B}$ are fixed in our study, and $0\le \beta_{A,B}\le 1$; for this interval $M_{A,B}'(0)$ are negative and a long-wave instability of the planar 
surface emerges, given $\alpha>0$.)  }
\label{table1}
\end{table}

\begin{figure}[h!]
%\vspace{-1.2cm}
\centering
\includegraphics[width=2.5in]{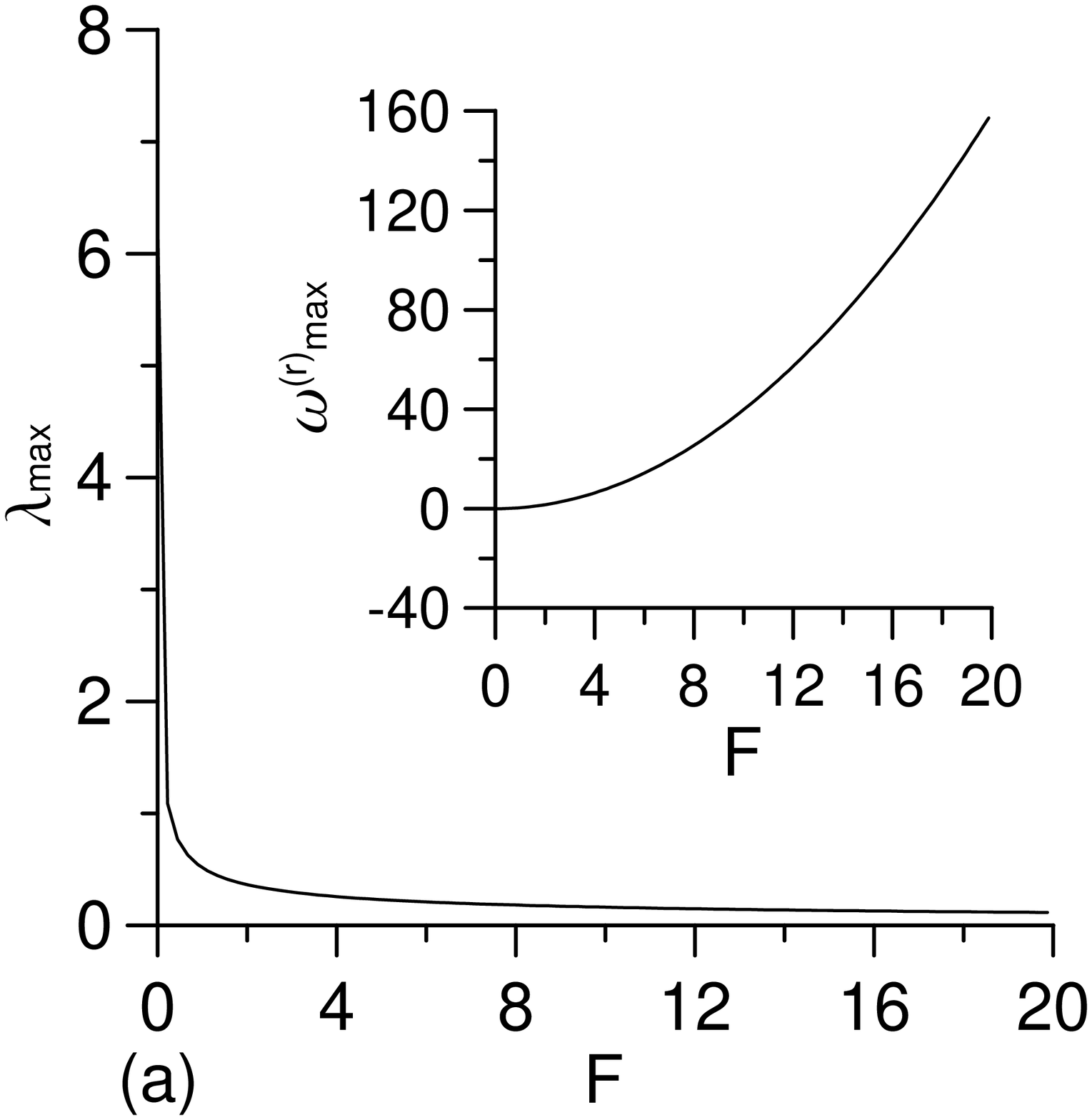}\includegraphics[width=2.5in]{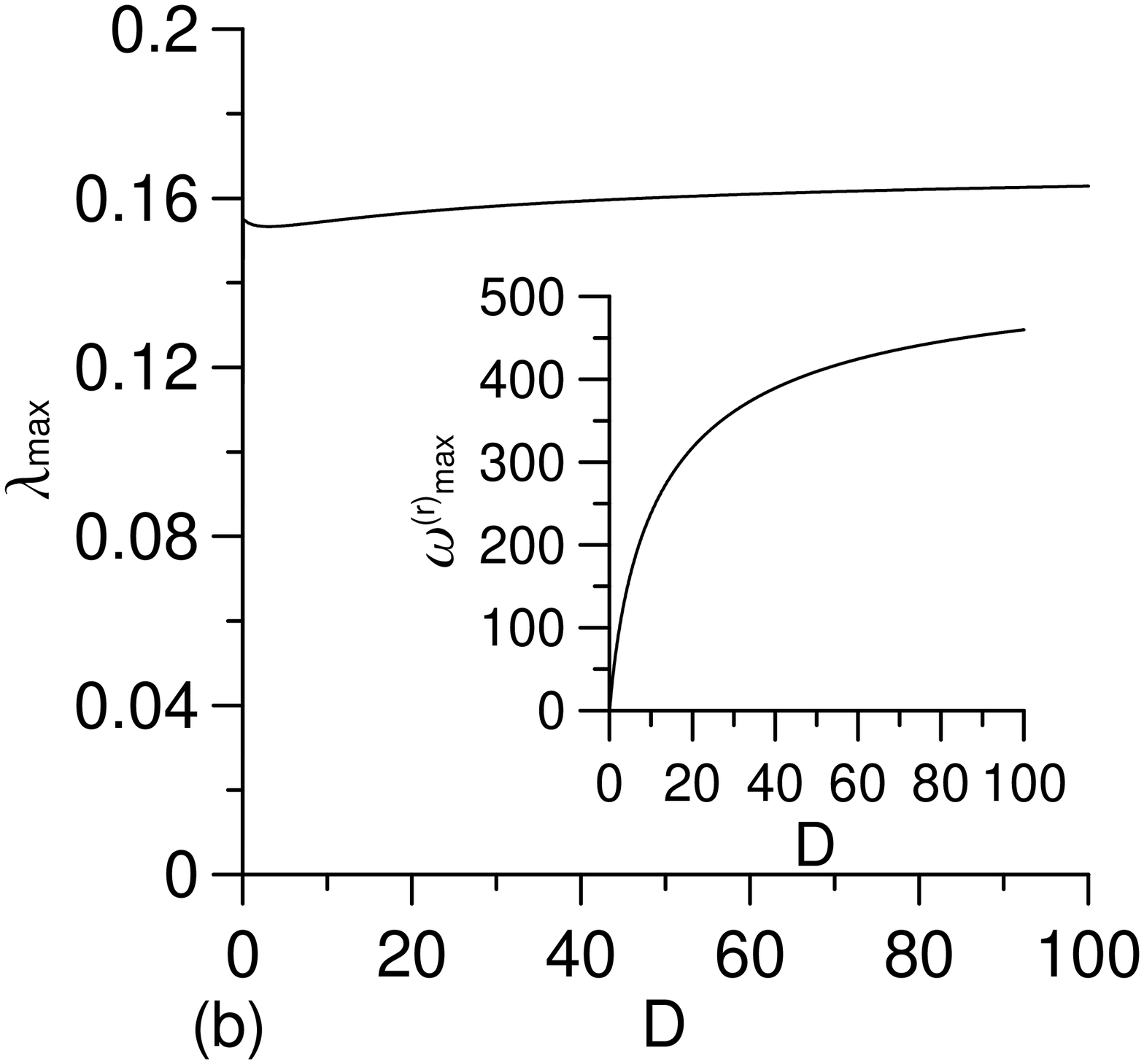}\\ \vspace{-1.2cm} \includegraphics[width=2.5in]{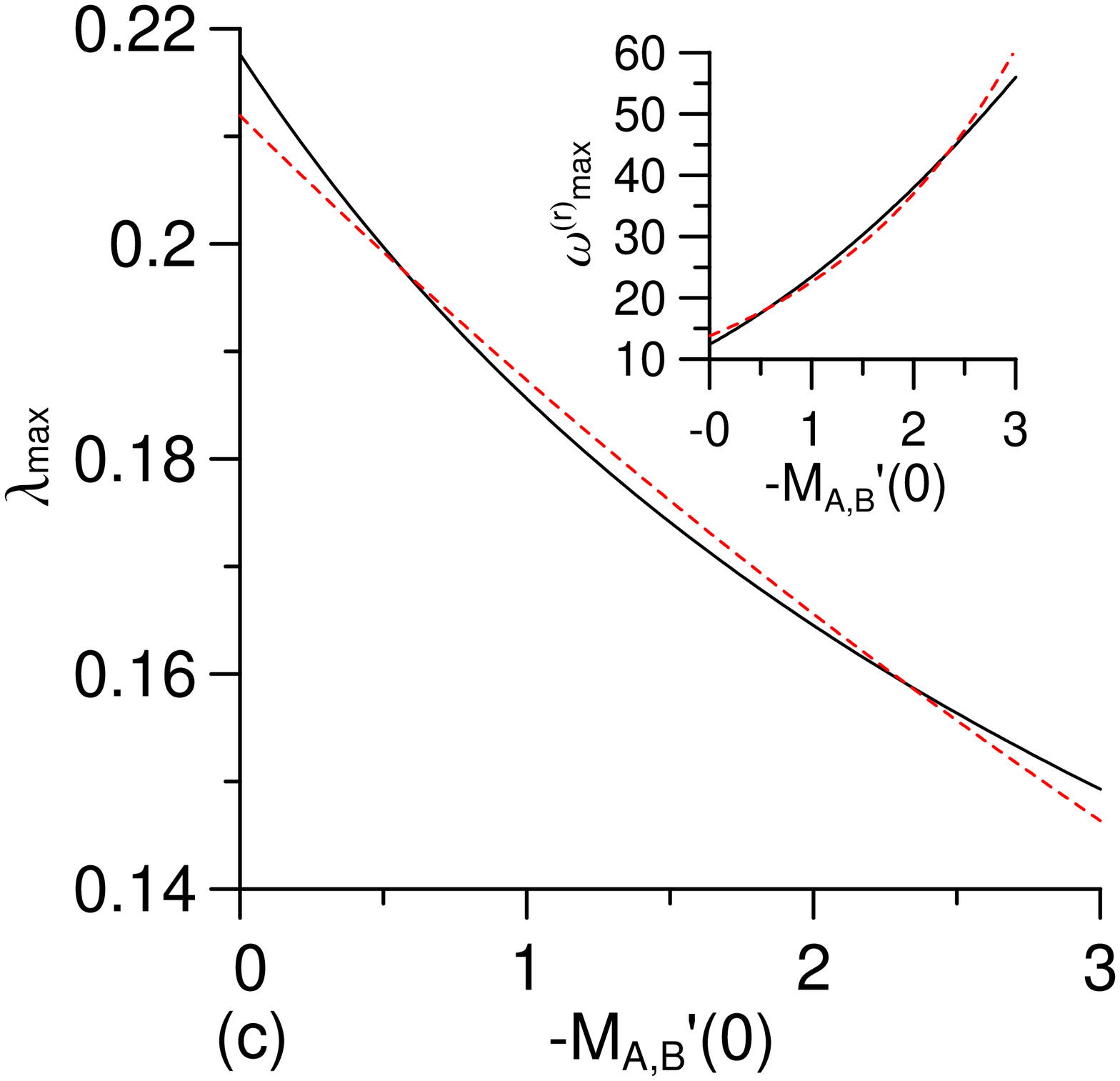}
\vspace{-0.8cm}
\caption{(Color online.) The most dangerous wavelength $\lambda_{max}$ and the corresponding growth rate $\omega^{(r)}_{max}$ vs. (a) the applied voltage parameter $F$, (b) 
the ratio of diffusivities $D$, and (c)
the derivatives of the diffusional mobilities 
evaluated at the planar unperturbed surface $h_x=0$, $M_A'(0)$ and $M_B'(0)$.  Dashed lines in (c): fits $\lambda_{max}=0.212\exp{(0.123M_{A,B}'(0))}$,  
$\omega^{(r)}_{max}=13.8\exp{(-0.493M_{A,B}'(0))}$.}
\label{lambda_omega_vs_FDMp}
\end{figure}

Other results of the LSA are shown in Figures \ref{lambda_omega_vs_FDMp}(a,b,c). Notice that we solved for $\lambda_{max}$ 
and $\omega^{(r)}_{max}$ numerically, since analytical solutions cannot be carried out by \textit{Mathematica} or \textit{Maple}. Increasing the applied voltage, 
the ratio of the diffusivities, and the absolute values of 
the derivatives of the diffusional mobilities result in the monotonic decrease of $\lambda_{max}$ and the increase of $\omega^{(r)}_{max}$, except that $\lambda_{max}$
very slowly increases with $D$. Interestingly, $k_c$ (and $\lambda_c$) do not depend on $D$, see Eq. (\ref{kc}).
The dependencies shown in Figure \ref{lambda_omega_vs_FDMp}(a) are very accurately fitted by the power laws $\lambda_{max}=0.514/\sqrt{F}$ and
$\omega^{(r)}_{max}=0.398F^2$ (the fits are indistinguishable from the curves in the figures), and those in Figure \ref{lambda_omega_vs_FDMp}(c) are fairly accurately fitted by the exponential functions
(see the caption to this Figure).  

Finally, from the condition that the determinant of the imaginary part of the system's matrix equals zero, it follows that
\[
\omega^{(i)}(k) = kF\left[ M_B(0) \left(1-\frac{m C_B^0}{4}\right)+\frac{M_A(0)}{4}D m C_B^0\right]. %\label{oi}
\]
Thus the perturbations also experience lateral drift
with the speed $v=|\omega^{(i)}(k)/k|$, which is proportional to the applied electric field parameter $F$ and does not depend on $k$.
Also $v$ increases linearly when the ratio of the diffusivities increases.  
%when one or both diffusional mobilities increase, or the ratio of the diffusivities increase. 
%Eq. (\ref{oi}) differs from the expressions derived by 
%Schimschak and Krug \cite{SK} and Bradley \cite{B} since in their local models the electric field is assumed constant on the surface.
%Using Eq. (\ref{oi}), the estimate of $v$ for the parameters in the Table \ref{table1} is 11.16, which translates into a dimensional value of 1.67 cm/s.
%Using fully nonlinear equations (\ref{nondim_h_eq2}) and (\ref{nondim_C_eq2}), from the computation of the evolution of a single perturbation wavelength 
%on a periodic domain 
%$0\le x\le \lambda_{max}$ (with the initial condition $\xi(x,0) = a\cos{k_{max}x},\ \hat C_B(x,0)=0$  or $\hat C_B(x,0) = b\cos{k_{max}x},\ \xi(x,0)=0$) we obtained a much smaller value 
%$v\sim 10^{-5}$ cm/s.
%
%\begin{figure}[h!]

In the next section we describe the computations of the surface morphology and composition evolution within the framework of the fully nonlinear system of PDEs (\ref{nondim_h_eq2}) and (\ref{nondim_C_eq2}). 

\setcounter{equation}{0}
\section{Nonlinear evolution of the surface morphology and composition}
\label{Computations}

Evolution equations (\ref{nondim_h_eq2}), (\ref{nondim_C_eq2}) are solved numerically using the method of lines \cite{MOL1,MOL2}. 
Integration in time is performed using the stiff ODE solvers RADAU \cite{RADAU} 
(implements a class of the implicit Runge-Kutta methods with automatic order switching)
and/or VODE \cite{DVODE} (implements a class of the backward differencing methods with automatic order switching), whereas the discretization in space is carried out in the conservative form using the second order finite differencing on a spatially uniform grid.

The computational domain is chosen $0\le x\le 20\lambda_{max}$, 
with the periodic boundary conditions for $h(x,t)$ and $C_B(x,t)$ at the endpoints of this interval. 
We tried two types of the initial conditions:  
$C_B(x,0)=1/2$, and a random, small-amplitude perturbation of the surface profile $h(x,0)=1$; or,
$h(x,0)=1$, and a random, small-amplitude perturbation of the concentration $C_B(x,0)=1/2$.
Evolution of the morphology and composition appears very similar for both initial conditions.

\begin{figure}[h!]
%\vspace{-1.2cm}
\centering
\includegraphics[width=2.5in]{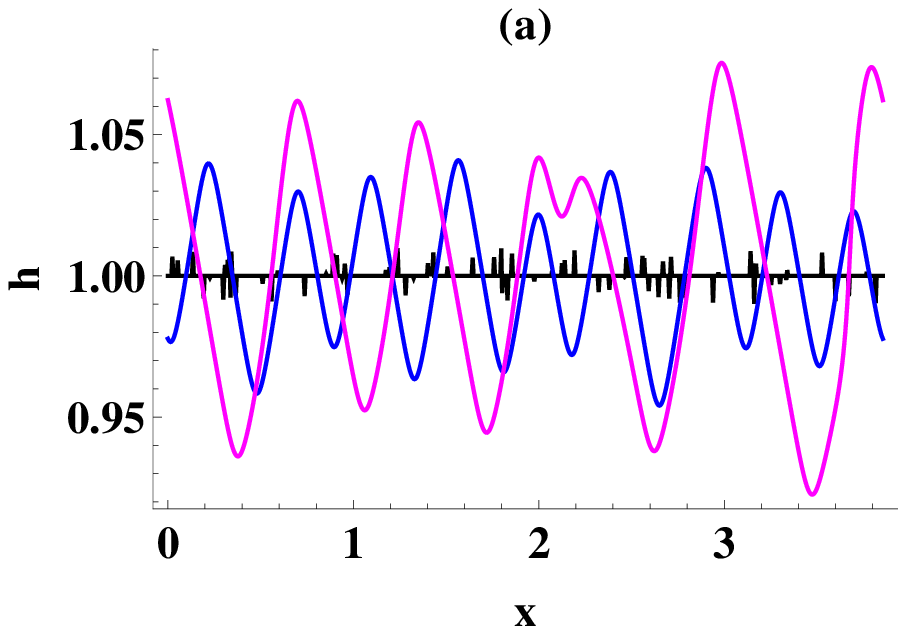}\hspace{1cm}\includegraphics[width=2.5in]{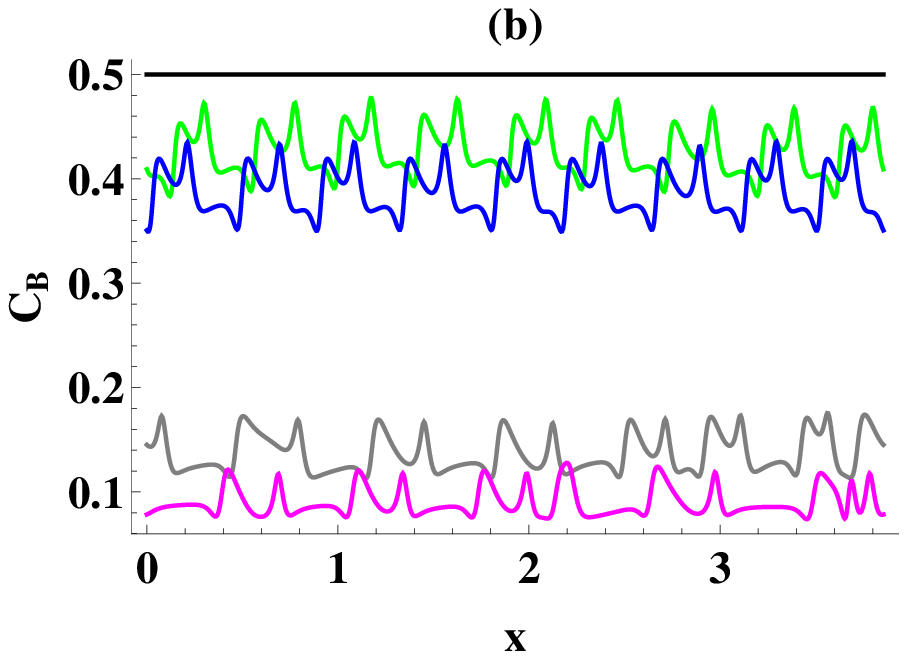}
%\vspace{-0.8cm}
\caption{(Color online.) Evolution of the morphology (a), and surface composition (b) from the initial condition $h(x,0)=1+\mbox{small random perturbation},\; C_B(x,0)=1/2$. $\beta_A=0.1,\; \beta_B=1$. The last profile (magenta line) corresponds to $t=10^{1.38}$, which is the last point on the dash-dot line in Figure \ref{ProfileMeanScale}(c). 
Same colors correspond to same $t$ value; many intermediate profiles are not shown.}
\label{h_CB_beta_A=0.1}
\end{figure}

\begin{figure}[h!]
%\vspace{-1.2cm}
\centering
\includegraphics[width=2.5in]{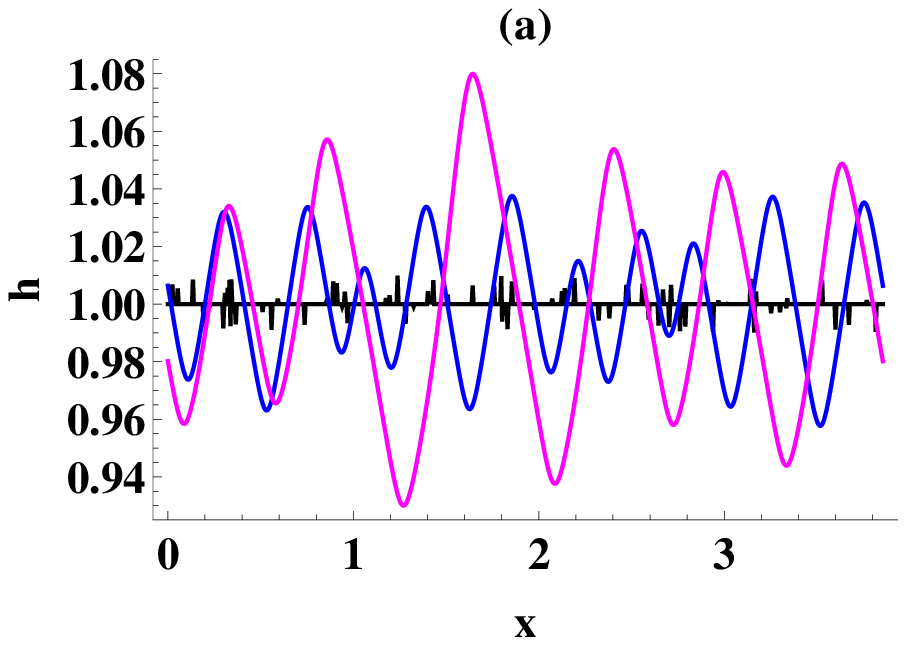}\hspace{1cm}\includegraphics[width=2.5in]{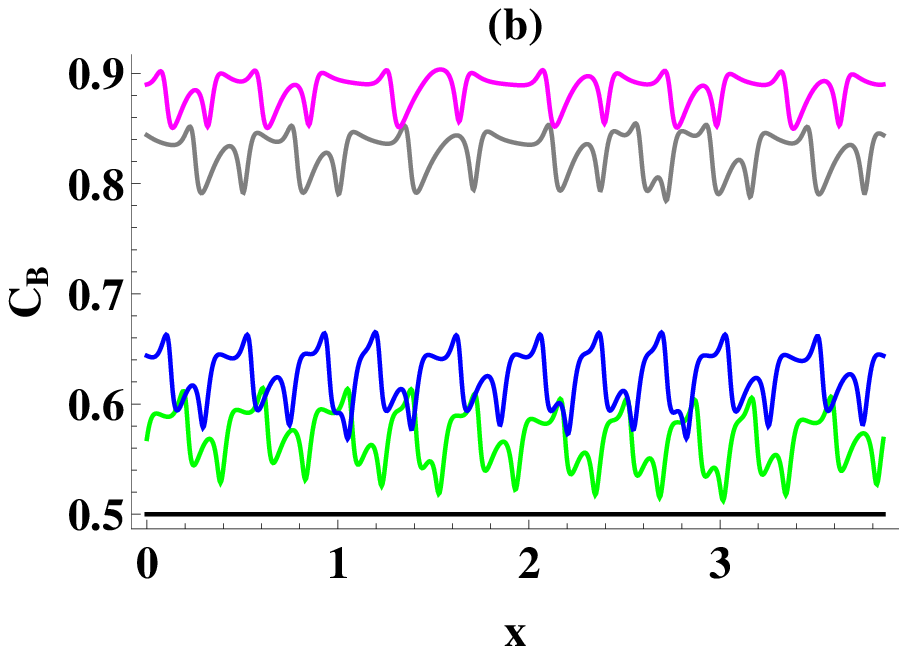}
%\vspace{-0.8cm}
\caption{(Color online.) (a), (b): Same as Figures \ref{h_CB_beta_A=0.1}(a,b), but $\beta_A=1,\; \beta_B=0.1$.}
\label{h_CB_beta_B=0.1}
\end{figure}

The surface morphology soon develops into a hill-and-valley structure,
and a perpetual coarsening of this structure sets in \cite{KD}-\cite{BMOPL},\cite{K} (Figures \ref{h_CB_beta_A=0.1}(a) and \ref{h_CB_beta_B=0.1}(a) show the examples). Unless the diffusivities or the diffusional mobility anisotropies of the two components 
significantly differ, i.e. $D\neq 1$ and/or $\beta_A \neq \beta_B$,
the concentration $C_B(x,t)$ relaxes fast to the mean value 1/2.
This means that when $D=1$ and $\beta_A = \beta_B$ and except during the aforementioned short relaxation period the morphology evolution can be described by the 
equation 
\[
h_t=\frac{2}{mQ}\frac{\partial}{\partial x}\left\{\left(1+h_x^2\right)^{-1/2}\left[\left(R\frac{\partial \kappa}{\partial x}
+F\right)\left(D M_A\left(h_x\right)+M_B\left(h_x\right)\right)\right]\right\}
%\label{nondim_h_eq2_late time}
\]
(notice that due to the choice of equal surface energies, see Table \ref{table1}, $R_A=R_B=R$). 

However, when $D\neq 1$ or $\beta_A \neq \beta_B$ the concentration $C_B(x,t)$ differs significantly from the mean value 1/2. When $D=0.1$ or $D=10$, $C_B$ fluctuates around
the mean value and instantaneous deviations are as large as 5\%. More interestingly, when $\beta_A=0.1\beta_B$ or vice versa, 
the evolution of the mean surface composition is markedly
different. In Figure \ref{h_CB_beta_A=0.1}(b), the mean value decreases, and the computation was terminated once $C_B$ reached zero locally. In contrast,
in Figure \ref{h_CB_beta_B=0.1}(b) the mean value increases and the computation was terminated when $C_B$ reached one locally. In the former case, the surface becomes enriched with
the component $A$, while in the latter case it is enriched with the component $B$. One can also notice that at the early and intermediate times the composition profiles in 
Figure \ref{h_CB_beta_B=0.1}(b) are nearly the mirror images of the ones in Figure \ref{h_CB_beta_A=0.1}(b); toward the end of the computation they are no longer.

\begin{figure}[h!]
%\vspace{-1.2cm}
\centering
\includegraphics[width=2.5in]{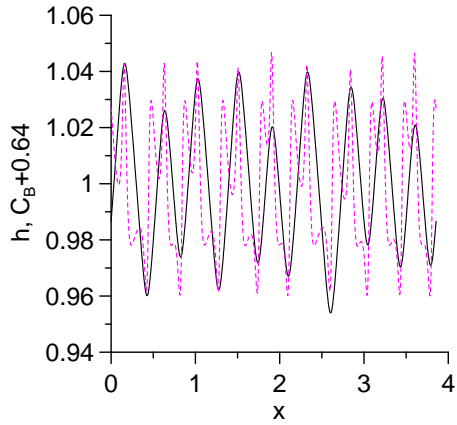}
\vspace{-1.0cm}
\caption{(Color online.) Concentration of $B$ adatoms (dashed line) superposed over the corresponding surface shape (solid line). The profiles are taken from Figures \ref{h_CB_beta_A=0.1}(a,b); 
see text for the discussion. }
\label{6th_H_CB_compare}
\end{figure}

In Figure \ref{6th_H_CB_compare} one of the computed $C_B$ profiles in Figure \ref{h_CB_beta_A=0.1}(b) is superposed
onto the corresponding surface shape from Figure \ref{h_CB_beta_A=0.1}(a). One can see that $C_B$ is the maximum (minimum) at the hill (valley), and in transitioning
from a hill to a valley (or vice versa) it behaves non-monotonically, that is, there is a local maximim (minimum) of $C_B$ on the downhill (uphill). 
So the hills (valleys) are richer in the component $B$ ($A$), and the hills slopes are alternatingly slightly richer in the $A$ and $B$ components. 
The difference in $C_B$ content of a hill and a valley is roughly 8\% in this Figure.


\begin{figure}[h!]
%\vspace{-1.2cm}
\centering
\includegraphics[width=2.5in]{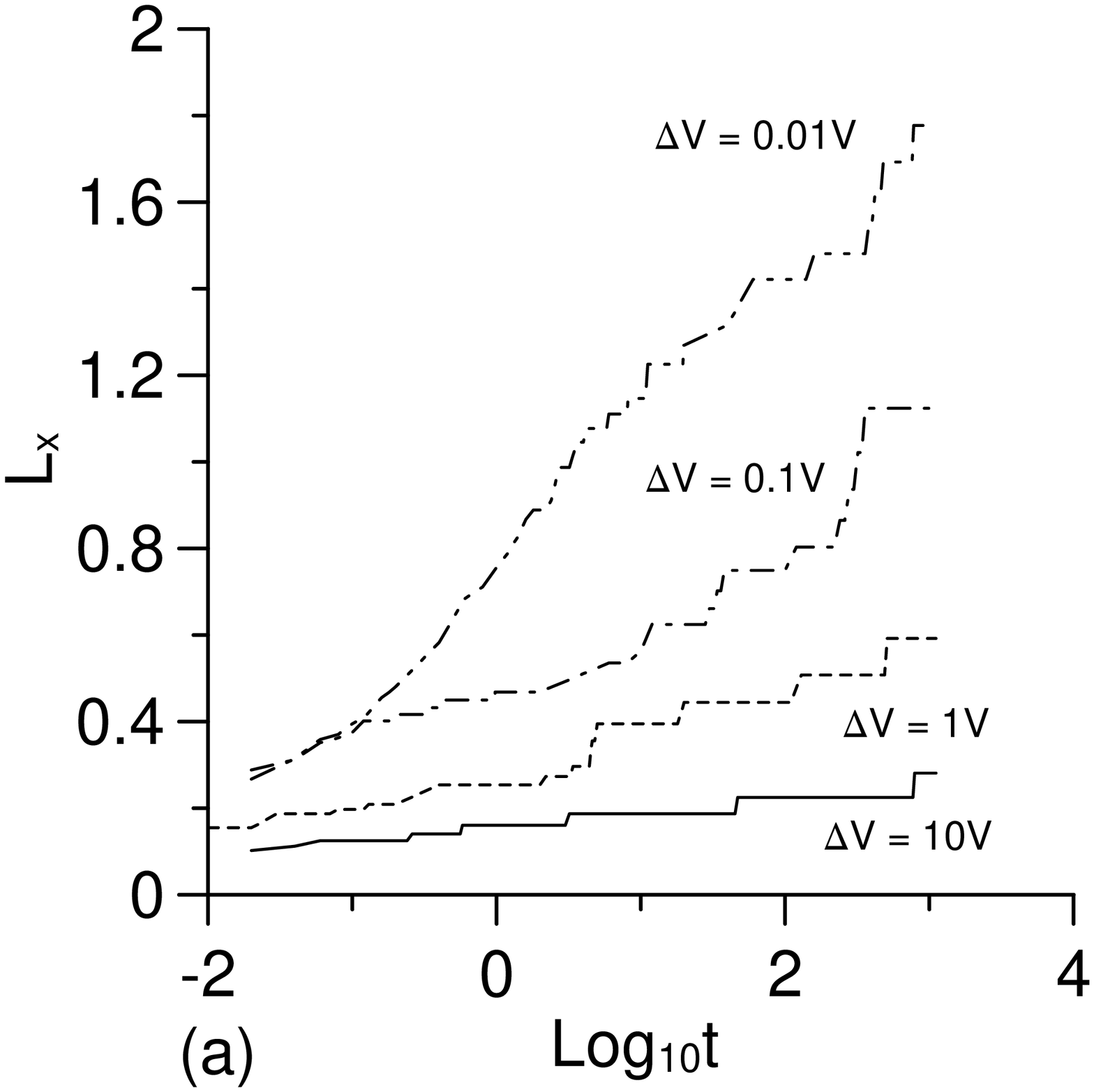}\includegraphics[width=2.5in]{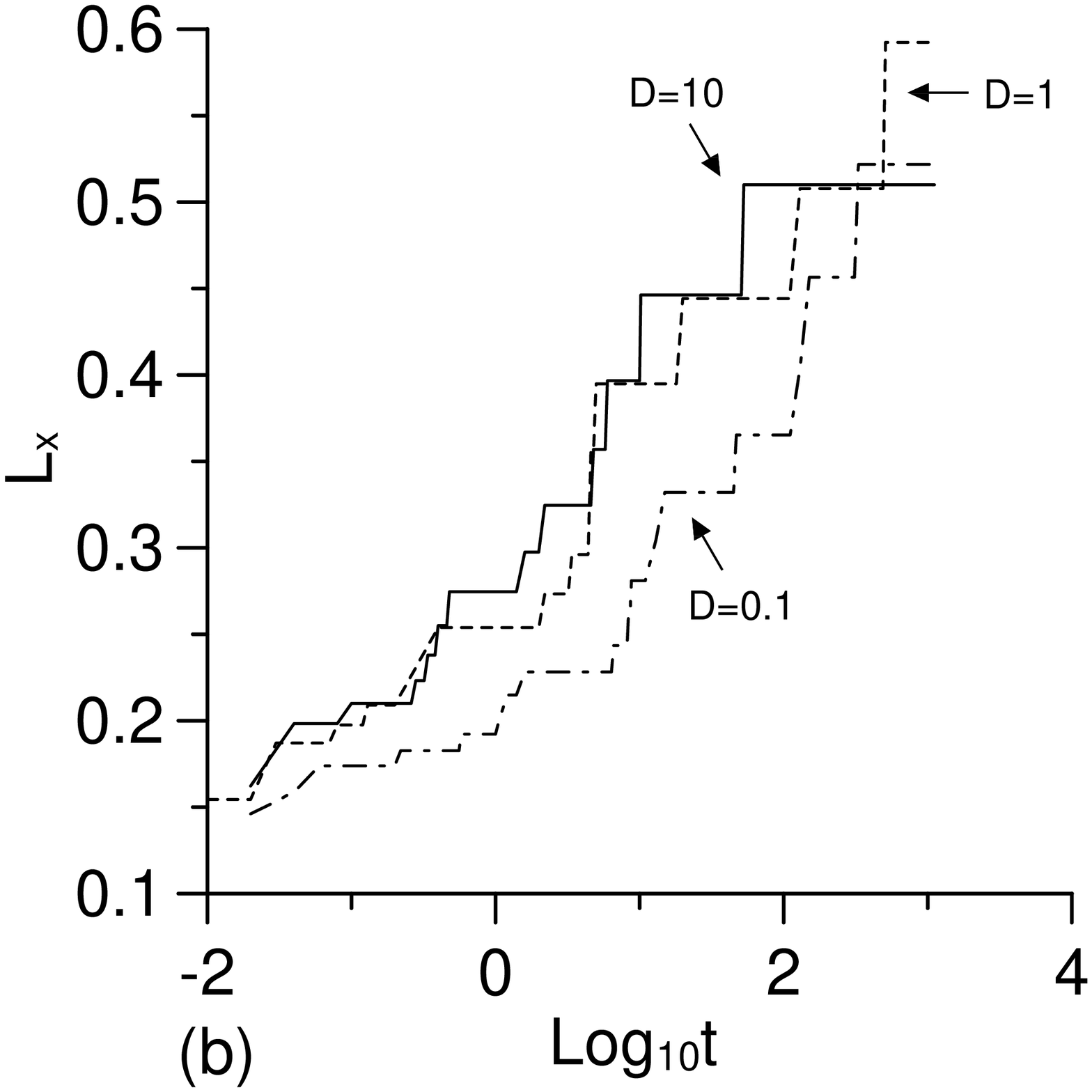}\\ \vspace{-1.2cm} \includegraphics[width=2.5in]{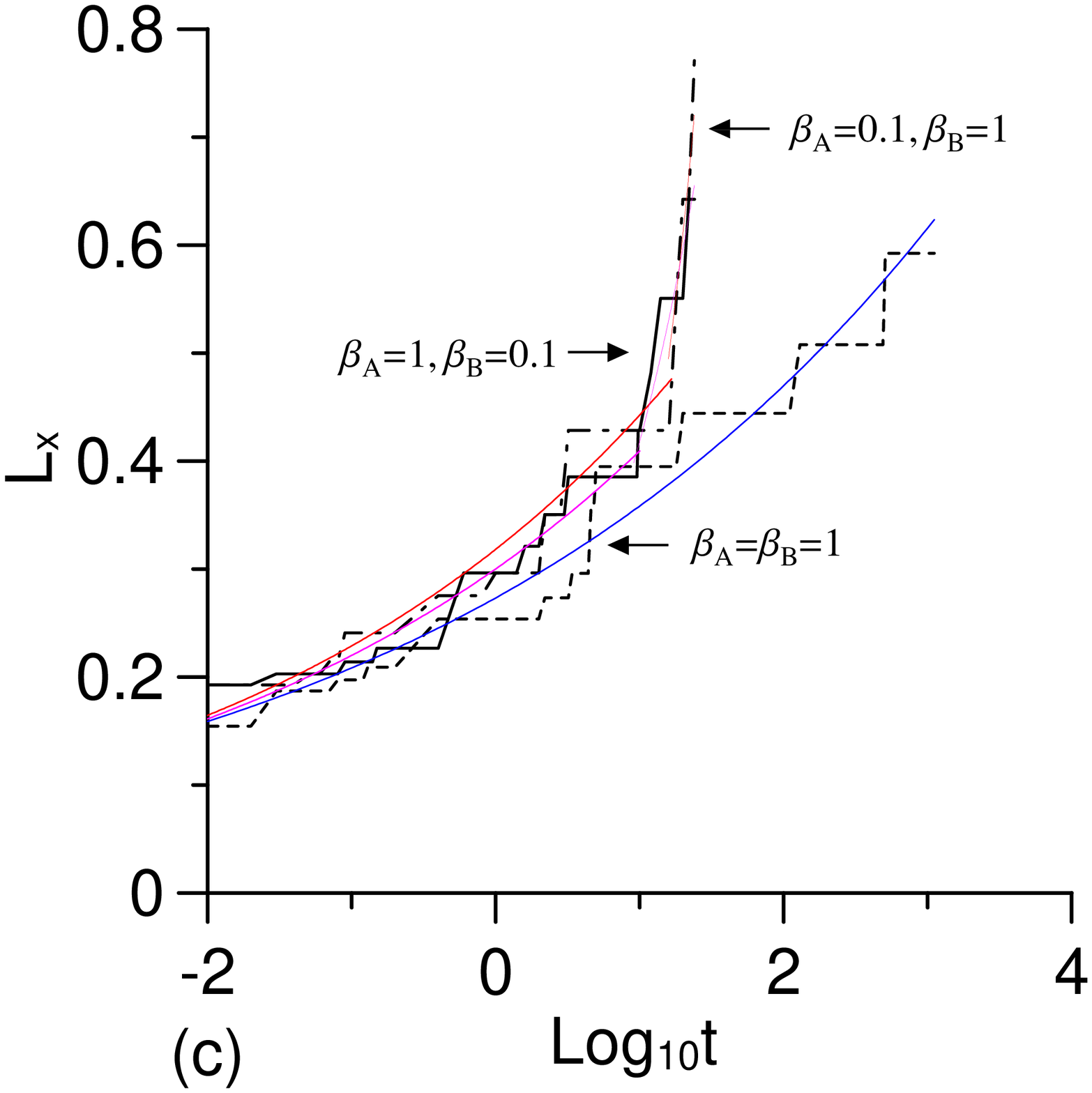}
\vspace{-0.8cm}
\caption{(Color online.) Evolution of the hill-and-valley structure's length scale. 
(a): Vs. $\Delta V$. For reference, $\Delta V=1V$ corresponds to $F=11.16$. (b): Vs. $D$. (c): Vs. $\beta_{A,B}$. The power law fits to the data are shown by thin solid lines (described in the text.)
}
\label{ProfileMeanScale}
\end{figure}

Next, we introduce the time-dependent characteristic lateral length scale of the hill-and-valley structure: $L_x= 20\lambda_{max}/(\mbox{number of valleys})$ (which has the meaning of the mean distance 
between the neighbor valleys), and
discuss how $L_x$ scales with the time and the key parameters (those marked as variable in Table \ref{table1}). 
Figure \ref{ProfileMeanScale}  has three panels, where the panel (a) shows how the time-dependence of $L_x$ scales with the applied voltage $\Delta V$; the panel (b) shows the scalings with $D$; and the panel (c) shows the scalings with $\beta_{A,B}$. When one parameter
is changed in a panel, all other parameters are fixed to their base values in the Table \ref{table1}.
Note that the flat horizontal segments of the curves correspond to the time intervals such that the hills slopes are slowly re-adjusting, which does not result in the changes
of the length scale; these changes occur when the slopes finally fall into the spinodal interval \cite{KD}.

\begin{itemize}
\item Variation of $\Delta V$ (Figure  \ref{ProfileMeanScale}(a)) 

Figure \ref{ProfileMeanScale}(a) shows that $L_x$ significantly decreases when $\Delta V$ increases, indicating that more hills and valleys 
fit into the computational domain at any given time, and thus, the coarsening of the surface morphology slows down as the electromigration intensifies. 
The natural surface diffusion attempts to planarize the surface, and the electromigration has the opposite effect of surface roughening, 
which is consistent with the LSA result that the most dangerous wavelength decreases with the increase of the applied voltage parameter.
The ratio of the final recorded length scales $L_x^{(\Delta V=0.01V)}/L_x^{(\Delta V=10V)}\approx 1.8/0.3=6$. 
Fitting gives the power laws coarsening $L_x^{(\Delta V=0.01V)}=0.77t^{0.138}$, $L_x^{(\Delta V=0.1V)}=0.46t^{0.126}$, $L_x^{(\Delta V=1V)}=0.273t^{0.118}$ and $L_x^{(\Delta V=10V)}=0.156t^{0.074}$. With the increase of the applied voltage from 0.01V  to 10V the exponent decreased nearly two-fold.

\item Variation of $D$ (Figure  \ref{ProfileMeanScale}(b))

In Figure \ref{ProfileMeanScale}(b), changing $D$ does not have very pronounced effect on coarsening. The power laws are $L_x^{(D=0.1)}=0.205t^{0.138}$, $L_x^{(D=1)}=0.273t^{0.118}$,
and $L_x^{(D=10)}=0.29t^{0.104}$.
Increasing $D$ results in slower coarsening.

\item Variation of $\beta_{A,B}$ (Figure \ref{ProfileMeanScale}(c))

Changing the anisotropies of the diffusional mobilities has the most drastic effect on coarsening rates. When the anisotropies are equal, the
single coarsening law $L_x^{(\beta_A=\beta_B=1)}=0.273t^{0.118}$ applies to the entire time interval. As pointed out above, in this case the concentration of $B$ adatoms 
stays close to the equilibrium value 1/2. When the anisotropies differ by a factor of ten, a pronounced speed-up of coarsening is observed. A single coarsening law
in these cases is inadequate: for the case $\beta_A=0.1, \beta_B=1$ we calculated $L_x^{(\beta_A=0.1,\beta_B=1)}=0.318t^{0.143}$ for $0\le Log_{10}t\le 1.22$ and 
$L_x^{(\beta_A=0.1,\beta_B=1)}=0.04t^{0.905}$ for $1.22\le Log_{10}t\le 1.38$; for the case $\beta_A=1, \beta_B=0.1$ we calculated $L_x^{(\beta_A=1,\beta_B=0.1)}=0.3t^{0.135}$ for $0\le Log_{10}t\le 1$ and 
$L_x^{(\beta_A=1,\beta_B=0.1)}=0.128t^{0.514}$ for $1\le Log_{10}t\le 1.38$. Thus there is a factor of 4-6 increase in the coarsening exponent, and in the former case 
the coarsening rate falls short of linear. The evolution of the surface composition for these cases is shown in Figures \ref{h_CB_beta_A=0.1}(b) and \ref{h_CB_beta_B=0.1}(b).

\end{itemize}

The final remark in this section concerns the linearization of the chemical potentials (Eq. (\ref{base_eq4})). 
To conduct the LSA, the potentials must be linearized, and we chose the linearization point $C_i=1/2$ as the convenient, ``neutral"  value which is typical for many binary alloys.
In other words, by means of the LSA and computation we studied the development of a spatially and temporally non-uniform surface composition from the initial state of 
equal concentrations of the alloy surface components (a sort of "phase separation"). Other linearization points are certainly possible, but their choice should be guided by 
the properties of a particular alloy.
%One can also consider a dilute alloy and linearize $\mu_A^0\left(C_A\right)$ about $C_A=c_a,\; c_a < 1/2$ and $\mu_B^0\left(C_B\right)$ about $C_B=1-c_a$, and then
%compute the dynamics.
Also, using the nonlinear chemical potentials instead of the linearized ones may affect the results of computations of the dynamics of the surface morphology and composition. However, we found that the only such effect is the slow-down of the dynamics as a whole, that is the stretching of the evolution time scale, but all coarsening 
exponents and evolution outcomes are unchanged.\footnote{When the full chemical potentials given by Eq.  (\ref{mu_c}) are factored into the derivation of the evolution PDEs, 
the result is that the term $\partial C_B/\partial x$ in Eqs. (\ref{nondim_h_eq2}) and (\ref{nondim_C_eq2}) is replaced by $\left(\partial C_B/\partial x\right)/\left(4C_B\left(1-C_B\right)\right)$.}

\setcounter{equation}{0}
\section{Conclusions}
\label{Conclusions}

We studied the electromigration-driven evolution of the surface morphology and composition for a bi-component solid film using a minimal and local model, 
which is nonetheless formulated as a coupled system of two heavily nonlinear parabolic PDEs. Both equations of this system reduce to the standard fourth-order surface evolution 
equation \cite{M} when the bi-component film is replaced by a single-component film, the electric field is turned off, and the diffusional mobility is isotropic. 

Through LSA and computation we established the parametric dependencies of the key quantities. Our results show the long-wavelength instability coupled to the lateral drift 
of the perturbations. The most dangerous wavelength $\lambda_{max}$ and the growth rate $\omega_{max}$ scale as $F^{-1/2}$ and $F^2$, respectively, where $F$ is the applied electric field parameter.
These scalings coincide with those obtained by Schimschak and Krug \cite{SK} using a nonlocal electric field model.
However, scalings of $\lambda_{max}$ and $\omega_{max}$ with the first derivative of the diffusional mobilities are different from Ref. \cite{SK}; there, they also scale as power law,
$\lambda_{max}\sim M'(0)^{-1/2}$, while in our model the dependence is closer to exponential. From the computations of surface morphology coarsening, we noticed 
that the characteristic exponents vary depending on which parameter is studied, but in most cases the exponents are at least two times smaller than the approximate 
value 1/4 reported by 
Krug and Dobbs \cite{KD} for the single-component film using the same local electric field model as ours. It is not clear from their paper whether this value is 
universal, i.e. does not depend on parameters.  The hills slopes are constant $\approx 38^\circ$ during coarsening, which is rather close to the $\approx 35^\circ$ 
reported by Krug and Dobbs. 
Finally, we noticed fast growth of the exponent in the terminal stages of coarsening with significantly different
values of the anisotropy strength in the expressions for the diffusional mobilities of the two components. In these cases the surface layer becomes nearly homogeneous 
at the end of the computation due to enrichment by either $A$ or $B$ component, thus the growth of the exponent is consistent with the previous statement.

\section*{Acknowledgements}
M.K. acknowledges support from the grant C-26/628 by the Perm Ministry of Education, Russia.

\end{document}